\documentclass[12pt]{iopart}
\usepackage{graphicx}  
\usepackage{bm}

\begin{document}

\title{Spectra, flow and HBT in Pb-Pb collisions at the LHC}

\author{P. Bo\.zek}

\address{Institute of Nuclear Physics,
PL-31342 Krak\'ow, Poland}

\address{Institute of Physics, Rzesz\'ow University, 
PL-35959 Rzesz\'ow, Poland}

\ead{piotr.bozek@ifj.edu.pl}

\begin{abstract}
The transverse momentum spectra,  elliptic flow and  interferometry radii
for Pb-Pb collisions at the LHC are calculated in  relativistic 
viscous hydrodynamics. For Glauber model initial conditions, we find that
 the data can be described  using a small value of shear viscosity 
$\eta/s=0.08$.
The viscosities and the equation of state  are the same as used for    RHIC 
energies.
\end{abstract}

The experiments with  Pb-Pb collisions at $\sqrt{s}=2.76$GeV at the LHC opened 
further  possibilities for the
 studies of the properties of the hot and dense matter. Compared 
to the highest RHIC energies the multiplicity of charged particles 
in central collisions increased by a factor $\simeq 2.4$ \cite{Aamodt:2010pb}.
The  fireball has a higher  energy density and lives longer than the one created in Au-Au collisions at $\sqrt{s}=200$GeV. 
At RHIC, the production of particles with soft momenta 
 can be described using the relativistic hydrodynamic model. A satisfactory description of the 
measured Hanbury Brown-Twiss  (HBT) radii requires the use of an equation 
of state  without a soft point at the transition from the quark-gluon plasma to the hadronic phase \cite{Broniowski:2008vp}. This observation is consistent with   lattice QCD calculations of the equation of state showing a crossover transition \cite{Aoki:2006we}.  Studies of 
the elliptic flow at different centralities of the collisions lead to the estimate of the ratio of the shear viscosity to entropy $\eta/s=0.08-0.2$
\cite{Luzum:2008cw}
,
 depending on  the assumed initial eccentricity of the source.

 \begin{figure}[thb]
 \includegraphics[width=0.44\linewidth]{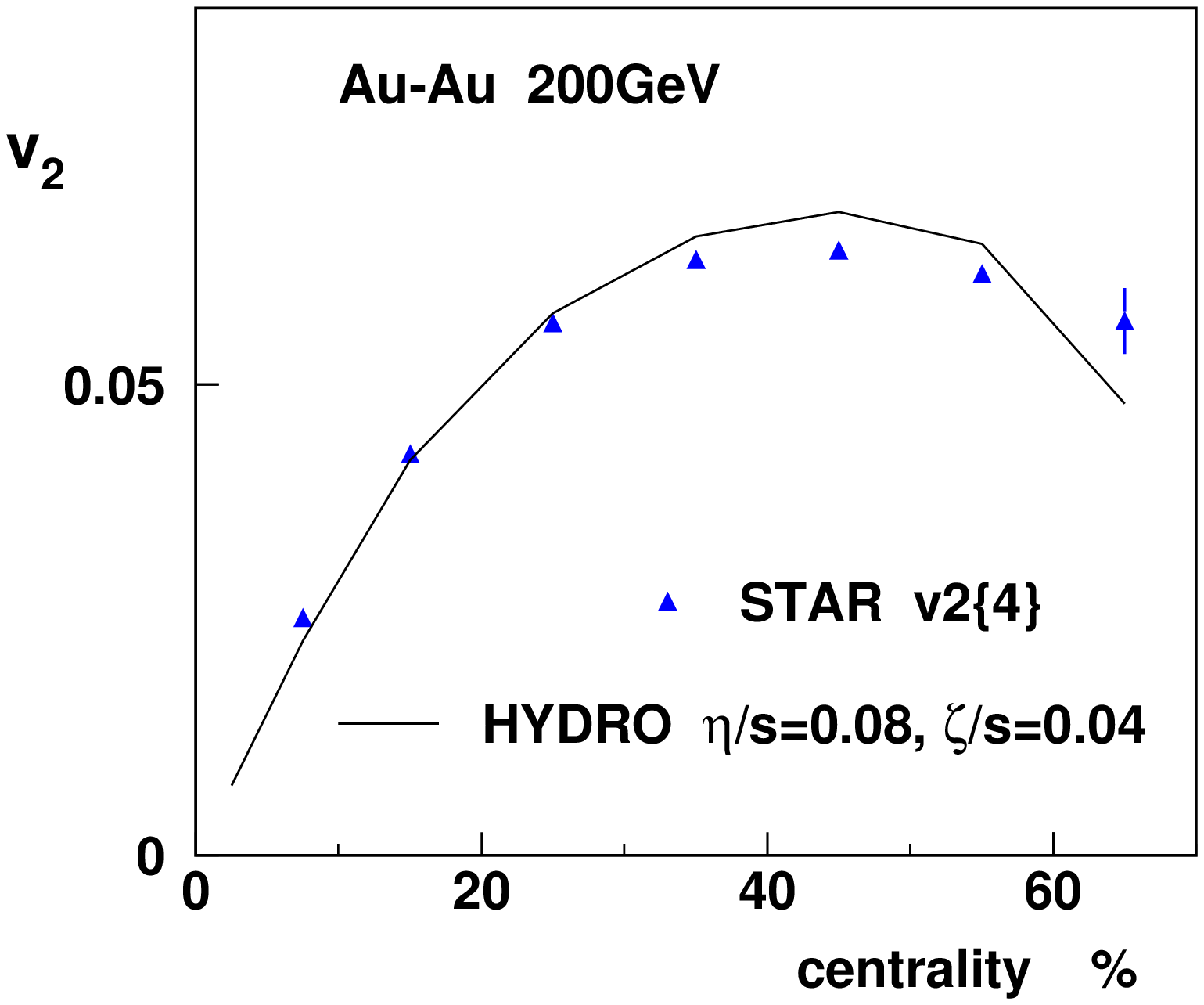}~~~~~
 \includegraphics[width=0.44\linewidth]{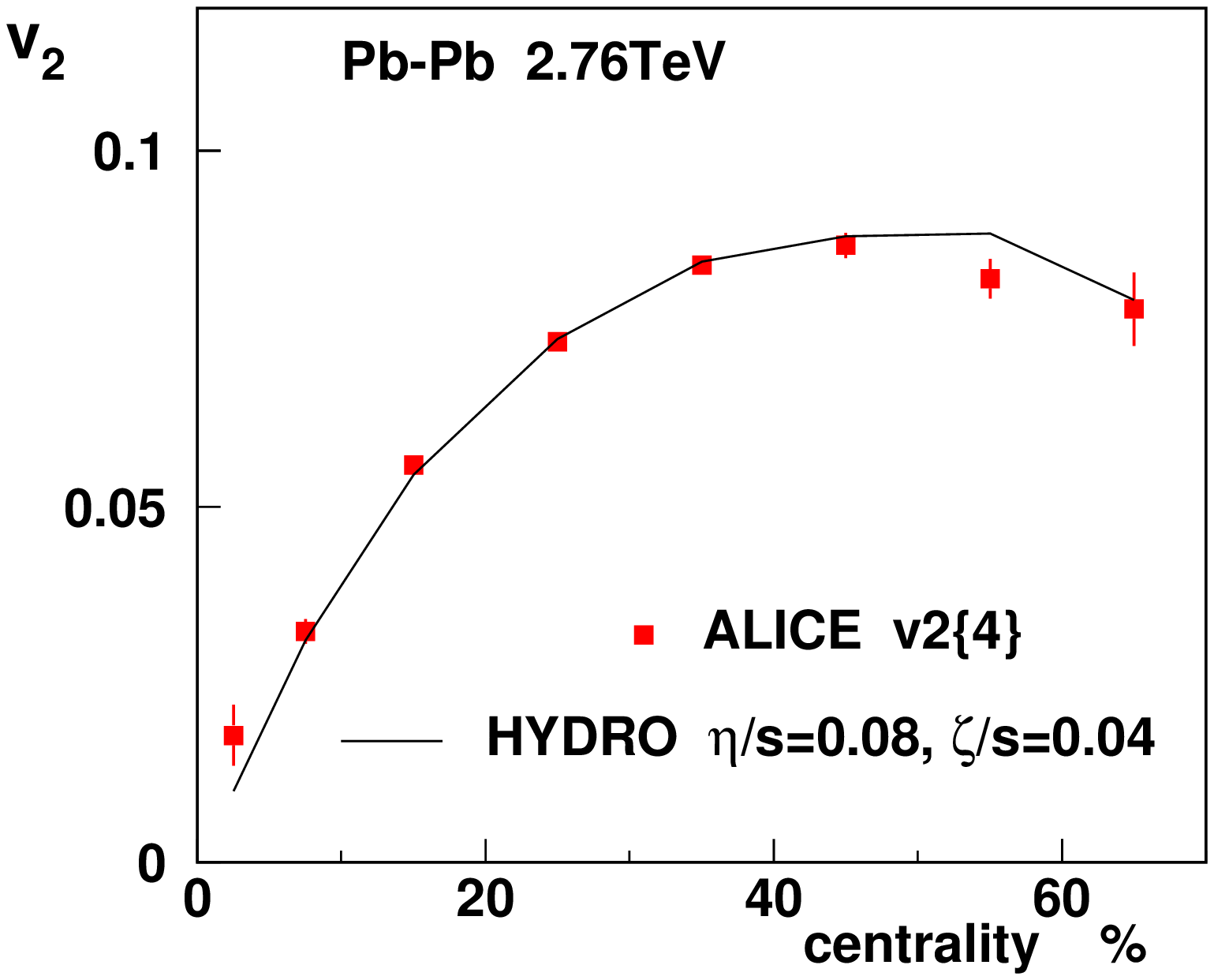}
 \caption{ (left panel) Elliptic flow coefficient in Au-Au collisions at $\sqrt{s}=200$GeV,
as function of centrality, STAR Collaboration data \cite{Adams:2004bi}. (right panel) same for Pb-Pb collisions at $2.76$TeV, 
 ALICE Collaboration data \cite{Aamodt:2010pa}.
 }
 \label{fig:v2ch}
 \end{figure}

 We apply  $2+1$ dimensional 
relativistic
viscous  hydrodynamics with shear and bulk viscosities 
to model the collision dynamics at $\sqrt{s}=2.76$GeV. 
The shear viscosity is fixed to the AdS/CFT value $\eta/s=0.08$, while the 
bulk viscosity is set to $\zeta/s=0.04$, in the hadronic phase. Second order viscous
 hydrodynamic equations \cite{IS} are solved and  hadrons are emitted 
at the freeze-out. The deviations from the equilibrium   
momentum distributions at freeze-out are implemented using a quadratic
   and an asymptotically  linear  ansatz 
in momentum for the shear
 and  bulk viscosity corrections respectively \cite{Bozek:2009dw}. 
An exponential ansatz 
for the bulk viscosity corrections leads to similar  results as
the linear one.
The initial entropy density  profile of the fireball is 
proportional to a  mixture of participant and binary collisions densities
\begin{equation*}
s(x,y) \propto {(1-\alpha)\rho_{part}(x,y,b)+2\alpha
 \rho_{bin}(x,y,b)} \ , 
\end{equation*} 
with  $\alpha=0.15$, fixed to  reproduce
 the measured centrality dependence of
 the charged particle multiplicity \cite{Bozek:2011wa}.

 \begin{figure}[thb]
 \includegraphics[width=0.51\linewidth]{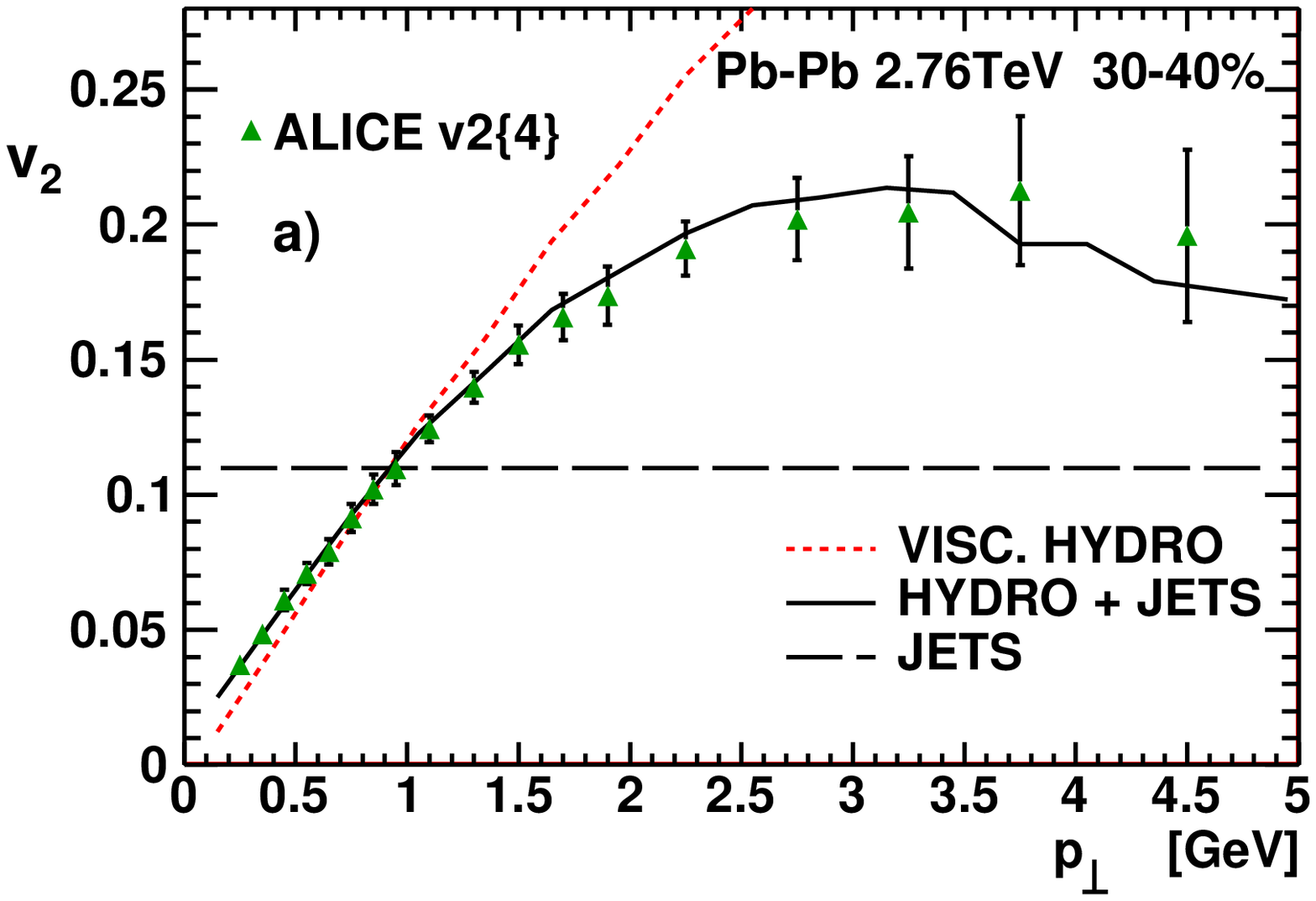}~~~~
 \includegraphics[width=0.51\linewidth]{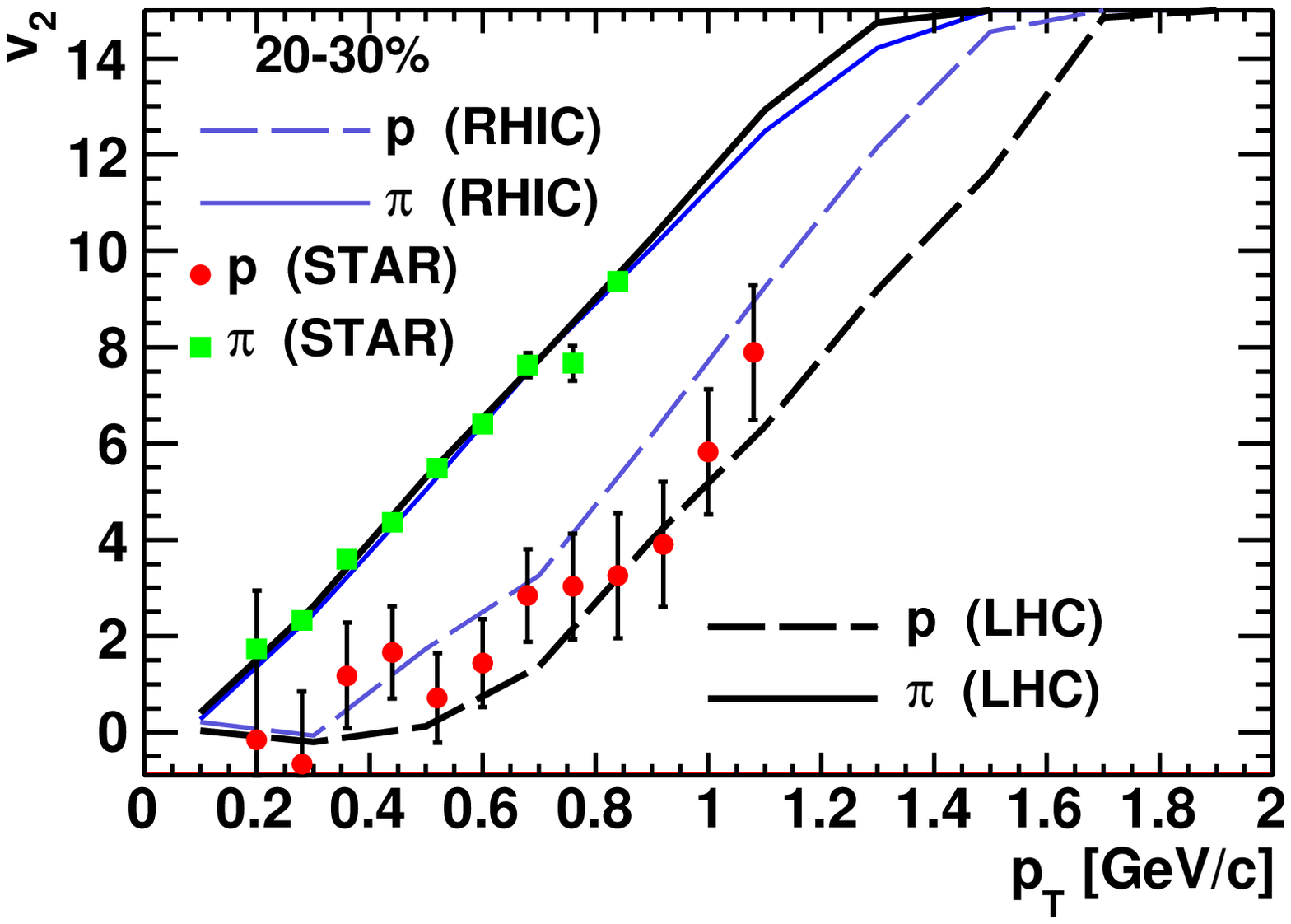}
 \caption{(left panel) Elliptic flow of charged particles
 as function of transverse momentum, ALICE Collaboration data \cite{Aamodt:2010pa}
compared 
to the results of relativistic hydrodynamics (dotted line) and to a schematic
 model including also a contribution from jets (solid line). (right panel)
The elliptic flow of identified particles at RHIC and 
at the LHC in relativistic viscous hydrodynamics,
  STAR Collaboration data  \cite{Adams:2004bi}.
 }
 \label{fig:v2pt}
 \end{figure}

The elliptic flow coefficient of charged particles $v_2$ is calculated 
for collisions at different centralities. 
It is significant that the elliptic flow at such very different collision  
energies can be described in a
satisfactory way using the same viscosity coefficients and the same equation of state  
(Fig \ref{fig:v2ch}). 
Assuming  Glauber model initial conditions, the shear viscosity coefficient 
that describes the data is small, $\eta/s\simeq 0.08$. 
 Other studies of the elliptic flow in heavy ion collisions at the LHC
show that the fluid has a small viscosity
\cite{Luzum:2010ag}
  $\eta/s=0.08-0.2$.
The elliptic flow created in central and semi-central collisions is well
described within viscious hydrodynamics with a constant value of $\eta/s$, 
there is no sign of a change of the shear viscosity to entropy ratio
 for higher temperatures
 reached at the LHC.

 \begin{figure}[thb]
 \includegraphics[width=0.55\linewidth]{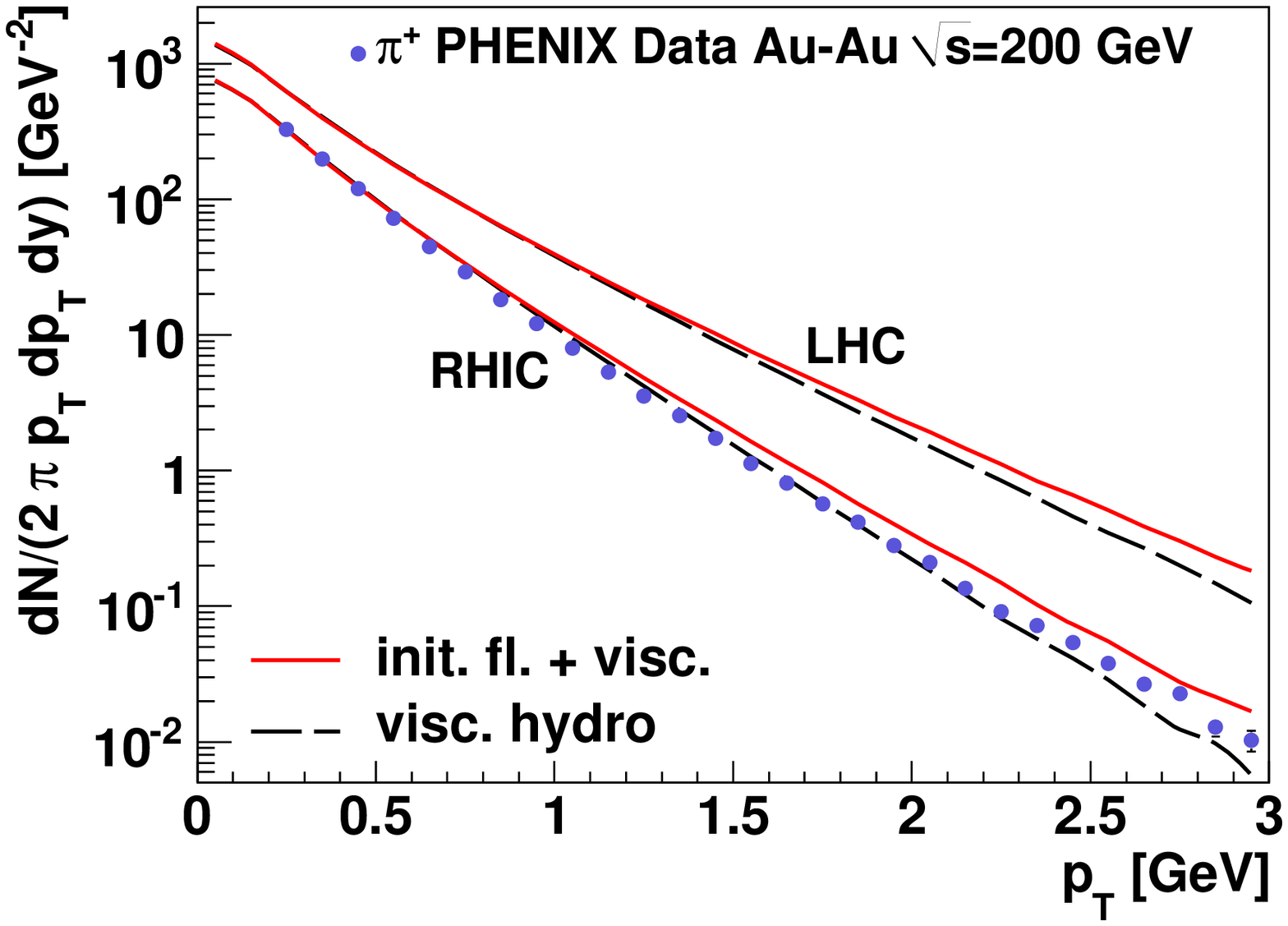}~~
 \includegraphics[width=0.38\linewidth]{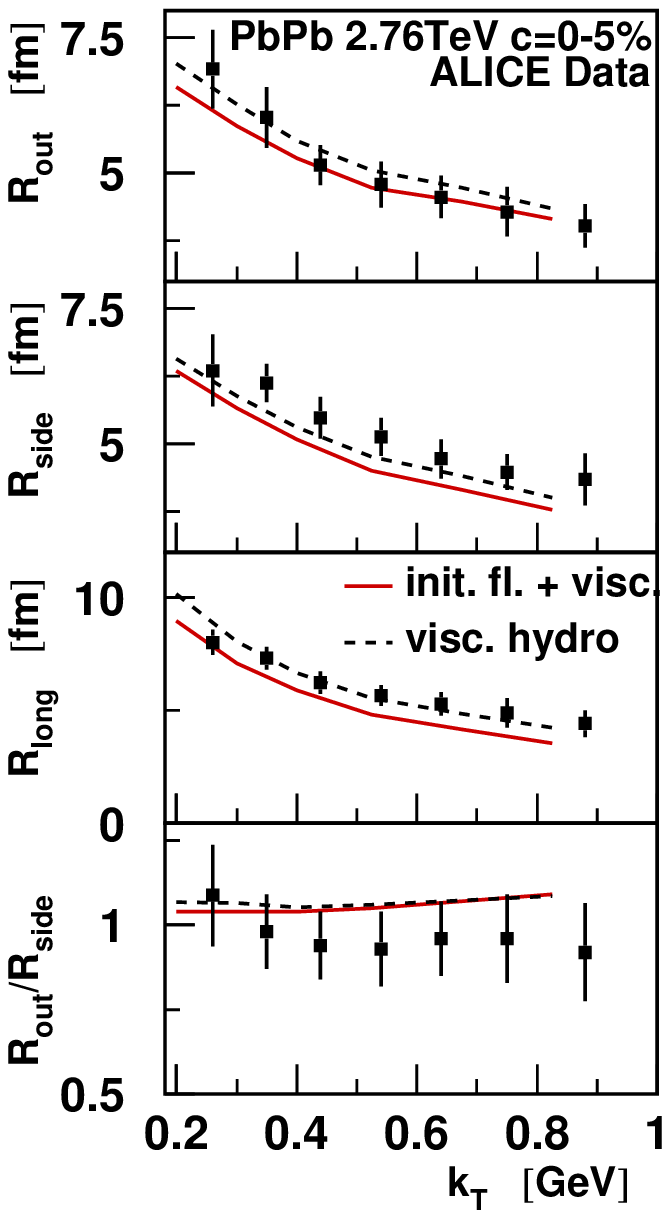}
 \caption{(left panel) Transverse momentum spectra of pions  at RHIC and the LHC, PHENIX Collaboration data \cite{Adler:2003kt}. (right panel)
The interferometry radii for pions emitted in Pb-Pb Collisions at
 $\sqrt{s}=2.76$GeV. The hydrodynamic model results are denoted  by the
 dashed lines for zero initial flow
 and by the solid lines when the initial pre-equilibrium flow is included 
\cite{Bozek:2010er},  ALICE Collaboration data
 \cite{Aamodt:2011mr}. }
 \label{fig:hbt}
 \end{figure}

 The measured  elliptic flow  as function of the transverse momentum 
shows systematic deviations from the predictions of the hydrodynamic 
model (Fig. \ref{fig:v2pt}, left panel). At high transverse momenta, 
such deviations are expected, as the majority of particles emitted at 
few GeV's originate from jets. 
 However, the data points at low momenta deviate from the 
hydrodynamic calculation as well. If this effect is not 
a systematic experimental 
bias, it could mean that a substantial contribution from non-thermalized 
remnants of jets is present; the elliptic flow from jets is 
of a  geometrical origin and could be larger than the one from 
the collective hydrodynamic flow at low momenta \cite{Bozek:2011wa}.
The elliptic flow of identified particles shows a stronger mass splitting
 at the highest energy. It is due to a stronger flow, which makes the 
corrections from bulk  viscosity more important. We note
 that in order to get  the mass splitting correctly  in a model
without a hadronic cascade after-burner a realistic non-zero bulk viscosity
 in the hadronic stage must be taken \cite{Bozek:2009dw}.

The strong transverse flow generated in the expansion of the hot fireball 
leads to flatter transverse momentum spectra at the LHC than at RHIC (Fig. \ref{fig:hbt},
 left panel)
\cite{Bozek:2010er}. The inclusion of the pre-equilibrium 
flow \cite{Vredevoogd:2008id}  makes the spectra harder, especially 
at RHIC energies. The HBT radii can be described to within $10-15$\%. The 
effect of the 
pre-equilibrium flow is small at $\sqrt{s}=2.76$TeV, 
 but improves  slightly the agreement with the data at $\sqrt{s}=200$GeV 
\cite{Bozek:2010er}. A similar quality in the data description  is
achieved in some earlier calculations using ideal fluid hydrodynamics \cite{Kisiel:2008ws}%
, especially when using modified profiles of the fireball or initial flow.
For collisions at the LHC, the hydrodynamic stage in the expansion dominates and the pre-equilibrium flow (if present) 
is relatively less important. We note that the preliminary transverse momentum spectra of pions, kaons and protons presented by the ALICE Collaboration
 at this conference can be described by the calculation without initial flow. 

We present the results of a viscous hydrodynamic model for 
Pb-Pb collisions at $\sqrt{s}=2.76$TeV. A  simultaneous
 description of the transverse momentum spectra, elliptic flow and 
HBT radii is achieved. It shows that relativistic hydrodynamics is a 
reasonable model of the expansion of the fireball and of the 
production of the 
bulk of the particles. The parameters of the model: the equation of state, the 
viscosity coefficients, the initial time and the 
freeze-out temperature are the same as 
deduced from the analysis of RHIC data. In particular, for Glauber model initial densities it means that $\eta/s=0.08$; an almost perfect fluid is produced in heavy-ion collisions at  the LHC.

\section*{Acknowledgement}

Supported by Polish Ministry of Science and Higher Education, grant N~N202~263438.

\end{document}